%% file: main.tex
\documentclass[12pt]{article}

\usepackage{amssymb,amsmath,amsfonts,eurosym,geometry,ulem,graphicx,caption,color,setspace,sectsty,comment,footmisc,caption,natbib,pdflscape,subfigure,array,hyperref,float}

\newcolumntype{L}[1]{>{\raggedright\let\newline\\arraybackslash\hspace{0pt}}m{#1}}
\newcolumntype{C}[1]{>{\centering\let\newline\\arraybackslash\hspace{0pt}}m{#1}}
\newcolumntype{R}[1]{>{\raggedleft\let\newline\\arraybackslash\hspace{0pt}}m{#1}}

\begin{document}

\begin{titlepage}
\title{Strategic Wealth Accumulation Under Transformative AI Expectations}
\author{Caleb Maresca\thanks{Email: ccm7752@nyu.edu. Department of Economics, New York University. Thanks to Philip Trammell, Basil Halperin, J. Zachary Mazlish, and Arden Berg for excellent comments and suggestions.\\
The code for this paper is available at \url{https://github.com/CalebMaresca/tai-capital-race}} \\
New York University}
\date{\today}
\maketitle
\begin{abstract}
\noindent This paper analyzes how expectations of Transformative AI (TAI) affect current economic behavior by introducing a novel mechanism where automation redirects labor income from workers to those controlling AI systems, with the share of automated labor controlled by each household depending on their wealth at the time of invention. Using a modified neoclassical growth model calibrated to contemporary AI timeline forecasts, I find that even moderate assumptions about wealth-based allocation of AI labor generate substantial increases in pre-TAI interest rates. Under baseline scenarios with proportional wealth-based allocation, one-year interest rates rise to 10-16\% compared to approximately 3\% without strategic competition. The model reveals a notable divergence between interest rates and capital rental rates, as households accept lower productive returns in exchange for the strategic value of wealth accumulation. These findings suggest that evolving beliefs about TAI could create significant upward pressure on interest rates well before any technological breakthrough occurs, with important implications for monetary policy and financial stability.\\
\vspace{0in}
\vspace{0in}

\bigskip
\end{abstract}
\setcounter{page}{0}
\thispagestyle{empty}
\end{titlepage}
\pagebreak \newpage

\doublespacing

\section{Introduction} \label{sec:introduction}

The accelerating pace of artificial intelligence (AI) development raises critical questions about its potential to reshape the global economy through two distinct yet interrelated mechanisms. First, AI systems capable of augmenting or replacing human researchers could dramatically accelerate scientific progress and economic growth, enabling parallel deployment of AI agents that rival human capabilities. Second, advanced AI—particularly artificial general intelligence (AGI)—could automate vast swaths of human labor, potentially concentrating economic benefits among capital owners while displacing workers. This looming possibility of automation may create novel incentives for strategic wealth accumulation today, as future control over AI labor could depend on wealth at the time of AI deployment. I term AI systems with these dual disruptive capacities Transformative AI (TAI), focusing on their specific economic implications.\footnote{This operationalizes \citet{GRUETZEMACHER2022102884}'s definition of Transformative AI as ``Any AI technology or application with potential to lead to practically irreversible change that is broad enough to impact most important aspects of life and society. One key indicator of this level of transformative change would be a pervasive increase in economic productivity.''}

To analyze how forward-looking economic agents adjust their current decisions in anticipation of TAI's uncertain arrival, this paper focuses specifically on the zero-sum nature of AI labor automation, distinct from AI's productivity-enhancing effects. When AI automates a job - whether a truck driver, lawyer, or researcher - the wages previously earned by the human worker don't vanish or automatically transform into broader economic gains. Instead, they flow to whoever controls the AI system performing that job. While AI will also generate new wealth through productivity gains (which this model captures through increased TFP growth), the reallocation of existing labor income creates immediate incentives for strategic capital accumulation.

To understand the full scope of TAI's economic implications, let us examine each of these transformative mechanisms in detail.

The prospect of AI accelerating scientific advancement is particularly compelling given that the number of scientific researchers appears to be a crucial factor driving economic growth \citep{jones2005growth}. If human-level AI is invented, many instances could be run in parallel, effectively multiplying the researcher population \citep{jones2022past}. Even without achieving human-level capabilities, AI systems could significantly enhance human researchers' productivity. Moreover, AI's ability to process and synthesize vast amounts of scientific literature could uncover connections that have eluded human scientists, who are necessarily limited in their capacity to absorb information\citep{agrawal2018finding}.

Turning to TAI's second major impact, its capacity for widespread automation raises important distributional concerns. Multiple leading AI developers explicitly pursue AGI systems that are ``generally smarter than humans,"\footnote{OpenAI: \url{https://web.archive.org/web/20250104180629/https://openai.com/about/} \ Google Deepmind: \url{https://web.archive.org/web/20250106123809/https://deepmind.google/about/}} which could render human labor economically obsolete across most domains. Unlike past technological disruptions that often created as many jobs as they eliminated, AGI could offer superior productivity across virtually all domains, potentially limiting the economic relevance of human labor. Unlike historical automation that created new roles, AGI might enable comprehensive substitution while concentrating returns: AI ``laborers" would generate wealth, but ownership of these systems would likely remain with existing capital holders. This represents not merely job displacement but a structural shift in labor’s role—from human activity to AI-mediated capital service, benefiting those with more capital at the expense of others.

Crucially, even uncertain TAI prospects could reshape present-day economic behavior. Households anticipating TAI may alter consumption, savings, and investment patterns years before it materializes. These forward-looking adjustments imply that expectations alone—independent of realized technological change—could generate significant macroeconomic effects today. Understanding this anticipatory channel is essential for policymakers and economists navigating AI’s economic implications.

This work extends \citet{chow2024transformative}, who model TAI expectations as either explosive growth or existential catastrophe, finding that short-term TAI forecasts elevate long-term interest rates via Euler equation dynamics. While retaining their focus on growth scenarios, I introduce two critical and interrelated innovations: (1) explicit modeling of labor reallocation from human workers to AI systems disproportionately owned by wealthy households, and (2) strategic interactions in savings behavior as households compete for future control over AI labor.

The redistribution mechanism creates novel economic dynamics. Households’ post-TAI labor supply depends on accumulated capital, incentivizing strategic savings to secure larger shares of AI-mediated production. Savings thus become both wealth-building tools and claims on future AI labor—a zero-sum competition absent in standard growth models. This disrupts traditional capital-pricing relationships: interest rates must now compensate not just for capital’s rental rate but for the expected value of AI labor control rights.

This competitive dynamic creates a form of prisoner's dilemma in savings behavior. While each household has an incentive to accumulate more wealth to secure a larger share of future AI labor, their collective actions to do so offset each other's relative gains. Everyone saves more, yet no one achieves the relative wealth advantage they sought. This leaves all households worse off through reduced consumption, even though the underlying pressure to accumulate wealth remains. This strategic mechanism, distinct from standard productivity growth effects, helps explain why interest rates can remain substantially elevated even as increased capital accumulation drives down productive returns.

My findings reveal that expectations of TAI can substantially affect current economic conditions, even before any technological breakthrough occurs. Under baseline scenarios with proportional wealth-based allocation of AI labor, I find one-year interest rates rising to 10-16\% compared to approximately 3\% without strategic competition, highlighting how anticipation of TAI can incentivize aggressive wealth accumulation. The effects strengthen as wealth becomes more important in determining future AI labor allocation, though with diminishing returns. Notably, interest rates diverge markedly from capital rental rates during the transition period - while increased savings drive down the marginal product of capital, interest rates remain elevated due to competition for future AI labor control. This wedge between productive returns and interest rates represents a novel channel through which technological expectations can influence financial markets. The magnitude of these effects varies with the assumed probability distribution of TAI arrival, with more concentrated near-term probabilities generating sharper initial increases in interest rates.

This paper proceeds as follows. \autoref{sec:model} details the model setup, with a focus on the novel mechanism for AI labor allocation. \autoref{sec:equilibria} characterizes equilibrium conditions, describes the solution method for computing transition paths from a pre-TAI economy to a post TAI-economy, and analyses the strategic wedge that develops between the capital rental rate and the interest rate during the transition process. \autoref{sec:result} presents quantitative results, analyzing how different assumptions about TAI arrival probabilities and wealth-sensitivity parameters affect interest rates and capital accumulation. \autoref{sec:conclusion} concludes with policy implications and directions for future research.

\section{Model} \label{sec:model}

This framework extends a neoclassical growth model to incorporate two fundamental effects of Transformative AI (TAI): (1) an acceleration of total factor productivity (TFP) growth, and (2) the displacement of human labor by AI labor. Households and firms operate under perfect foresight except for uncertainty about when TAI will be invented, which follows an exogenous arrival process.

\subsection{Households and Firms} \label{subsec:households}

A continuum of measure 1 households maximizes expected lifetime utility with constant relative risk aversion (CRRA) preferences:

\begin{equation*}
    \max_{\{c_t\}_0^\infty} \mathbb E \sum_{t=0}^{\infty} \beta^t \frac{c_t^{1-\eta} - 1}{1 - \eta}
\end{equation*}
\begin{equation*}
    \begin{aligned}
    \text{s.t.} \quad k_{i,t+1} + b_{i, t+1} + c_{i,t} & \leq w_t l_i + (1 + r_{k,t}) k_{i,t} + (1 + r_{b,t})b_{i,t} \\
    c_{i,t} \geq 0 & \text{ \& } k_{i,t} \geq 0 \quad \forall t\geq 0,
    \end{aligned}
\end{equation*}

\noindent where $ c_{i,t}$, $ k_{i,t}$, $b_{i,t}$, and $ l_i $ are household i's time $t$ consumption, time $t$ capital claims, time $t$ bonds, and labor supply, respectively. $ w_t $ is the time $t$ wage rate. $r_{b,t}$ and $ r_{k,t} $ are the time $t$ interest rate and time $t$ capital rental rate, respectively. $\beta$ is the discount rate, and $\eta$ is the coefficient of relative risk aversion. The expectation operator $\mathbb{E}$ accounts for uncertainty over TAI’s invention timeline. I differentiate between capital and bonds instead of lumping them togeather as ``assets" because a key part of my analysis will show how the effects of strategic savings will create a wedge between the capital rental rate and the interest rate.

A representative firm produces output via Cobb-Douglas technology:

\begin{equation*}
    Y_t = K_t^\alpha (A_t L)^{1 - \alpha},
\end{equation*}

\noindent where $ \alpha $ is the capital share, $ K_t $ is aggregate capital, $ L $ is aggregate labor, and $ A_t $ is total factor productivity (TFP). TFP is assumed to grow at an exogenous rate of $g_{SQ}$ (status-quo) before TAI and $g_{TAI}$ thereafter. The firm maximizes profits according to the following problem:

\begin{equation*}
    \max_{K_t,N} \{ K_t^\alpha (A_t N)^{1 - \alpha} - (r_{k,t} + \delta)K - w_tL  \}
\end{equation*}

\noindent where $\delta$ is the depreciation rate. 

\subsection{Invention of TAI and its effects} \label{subsec:tai-effects}

Households share homogeneous beliefs about TAI’s invention timeline, represented by an exogenous probability distribution over arrival dates. This distribution may sum to less than 1, allowing for the possibility that TAI never materializes.\footnote{This does not necessarily correspond with the event that AGI is never invented. Perhaps AGI is invented, but it does not radically alter the economy.}

Upon TAI’s invention, two immediate regime shifts occur. First, productivity growth rises permanently from $g_{SQ}$ to a $g_{TAI}$. Second, human labor is fully replaced by AI labor. Aggregate labor supply $N$ remains constant at its pre-TAI level, but its composition shifts from human workers to AI labor units. Crucially, these units are reallocated across households based on wealth at the time of TAI arrival. This design choice serves to isolate redistribution effects: by holding aggregate labor constant, I neutralize scale effects and ensure that all economic growth post-TAI is attributable to the increase in the growth rate of TFP.

In the competition over the control of AI labor, it is assumed that agents who are richer at the time TAI is invented will capture a larger fraction of the wealth created by TAI. This could occur directly through market mechanisms or through political processes. I abstract from the exact mechanism through which this conflict and reallocation occur and instead provide a simple formula by which AI labor is allocated asymmetrically toward the rich. Specifically, while $l_i$ was constant across households prior to TAI, after TAI $l_i$ is reallocated according to the following formula:

\begin{equation} \label{eq:li}
    l_i(k_{i,t_{TAI}},K_{t_{TAI}}) = \frac{\left(\frac{k_{i,t_{TAI}}}{K_{t_{TAI}}}\right)^\lambda}{\int_0^1 \left(\frac{k_{i,t_{TAI}}}{K_{t_{TAI}}}\right)^\lambda di}  \text{ for } \lambda \in \mathbb{R},
\end{equation}

\noindent where $k_{i,t_{TAI}}$ is the capital claims of the individual at the time TAI is invented, $K_{t_{TAI}}$ is the aggregate capital at that time\footnote{which is also equal to the average among the households, as there are measure one households, see the market clearing equations in \autoref{subsec:markets}.}, and $\lambda$ is a parameter that determines how sensitive the allocation of AI labor is to the wealth of the individual.

The allocation mechanism in equation (1) captures how control over automated labor becomes a zero-sum competition. Unlike traditional capital which can be produced to meet demand, the total supply of labor (human or AI) remains fixed in the model. When AI automates a task, it doesn't create new labor - it redirects existing labor returns from human workers to AI owners. Any productivity enhancements from AI are captured separately through the increase in TFP growth rate from $g_{SQ}$ to $g_{TAI}$. This modeling choice isolates the redistributive effects of automation from its productivity effects.

An alternative approach would be to model AI labor as a stock that can be accumulated through investment, similar to physical capital. This could better capture the possibility of expanding AI deployment over time. However, such an extension would significantly complicate the model's dynamics by introducing a second type of capital with its own accumulation process and strategic implications. I leave this promising extension to future research, focusing here on the core mechanism of wealth-based allocation of a fixed labor supply.

Intuitively, this equation allocates AI labor to each individual based on their relative wealth at the time of TAI invention, with the parameter $\lambda$ determining how strongly wealth differences translate into differences in AI labor allocation. When $\lambda = 0$, AI labor is distributed equally across the population regardless of wealth, implying no reallocation of labor. Conceptually, one can imagine that all human labor is automated, but each household is given an AI laborer (by the government or some other redistributing agency) exactly replacing their labor and leaving their economic situation unchanged. Equivalently, one can imagine that this corresponds to the outcome where TAI increases the growth rate but does not automate human labor. Therefore, this case is equivalent to \citet{chow2024transformative}\footnote{Ignoring the possibility of existential catastrophe.} and will be useful for comparison.

When $\lambda > 0$, wealthier individuals receive disproportionately more AI labor, with higher values of $\lambda$ leading to a concentration of AI labor among the wealthy. $\lambda = 1$ corresponds to the case where AI labor is allocated proportionally to wealth—a household that is 10\% richer when TAI is invented will be allocated 10\% more AI labor. Higher values of $\lambda$ intensify strategic competition at the top of the wealth distribution while diminishing the stakes for those at the bottom. For instance, with high $\lambda$, small wealth differences between the richest households translate into large differences in AI labor allocation, while similar wealth gaps between poorer households have minimal effect. This parameter can thus capture various intensities of wealth-based competition, from purely proportional allocation ($\lambda = 1$) to nearly winner-take-all scenarios ($\lambda \to \infty$). There is good reason to believe that $\lambda$ may be significantly greater than one, as control over AI systems may concentrate disproportionately among the wealthiest actors.

One can also consider $\lambda < 0$, whereby AI labor is distributed disproportionately toward poorer individuals. However, I do not focus on this case as I assume that wealth provides advantages in securing AI resources.

\subsection{TAI Timelines}

Household beliefs regarding the year in which TAI will be invented are treated as exogenous. While this assumption simplifies the analysis, it abstracts from plausible feedback mechanisms. For instance, the pace of AI development likely depends on endogenous factors such as private R\&D investment and policy interventions, which could correlate with household expectations—especially under high values of $\lambda$, where wealthier agents anticipate disproportionate gains from automation. Nevertheless, the exogenous timeline assumption provides a tractable foundation for isolating the economic effects of belief-driven behavior.

TAI invention is modeled as a stochastic process requiring $n$ technical breakthroughs, each with heterogeneous difficulty. This motivates using a negative beta binomial distribution (NBN), which generalizes the negative binomial by allowing success probabilities to follow a beta distribution.\footnote{Following the convention where the negative binomial counts total trials until $n$ successes, not failures.} Each trial represents a potential breakthrough in a given month. These monthly probabilities are then aggregated to obtain annual probabilities. 

To capture uncertainty about the number of breakthroughs required for TAI, I treat $n$ as a discrete random variable drawn from a bounded support $\{ n_{\text{min}}, ..., n_{\text{max}} \}$. The resulting compound distribution integrates uncertainty over both breakthrough requirements and success probabilities. This distribution is truncated at 60 years for computational feasibility, with residual probability mass reallocated to the event that TAI is never invented.

Households update timeline probabilities annually via Bayesian filtering. For tractability, current results focus on passive learning, wherein probability mass shifts from elapsed years to remaining possibilities as time progresses without TAI being invented. A possible extension could involve active learning, wherein households would update their posterior beliefs in accordance with the number of observed breakthroughs each year.

\subsection{Calibration}

\begin{figure}[h]
    \centering
    \includegraphics[width=0.85\textwidth]{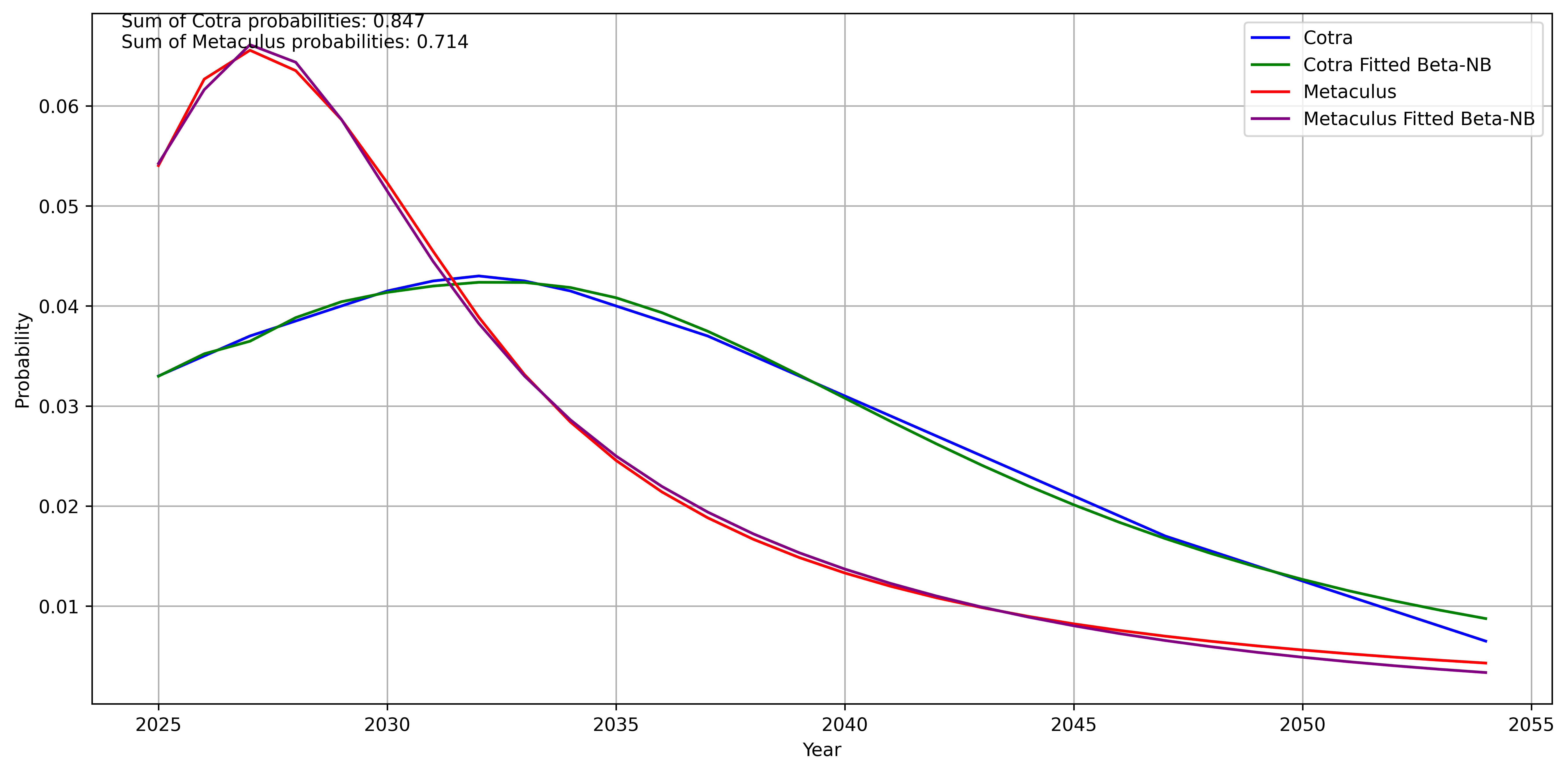}
    \caption{ \small Predicted Probabilities of TAI Arrival. The figure shows fitted negative beta-binomial distributions based on forecasts from Cotra (2023) and Metaculus predictions.}
    \label{fig:probabilities}
\end{figure}

The yearly TAI probabilities are calibrated using two primary sources of AI timeline estimates: forecasts by Ajeya Cotra,\footnote{I selected yearly probabilities to roughly align with \url{https://www.alignmentforum.org/posts/AfH2oPHCApdKicM4m/two-year-update-on-my-personal-ai-timelines} and \url{https://www.alignmentforum.org/posts/K2D45BNxnZjdpSX2j/ai-timelines}} a Senior Advisor at Open Philanthropy who has conducted extensive research on AI development timelines, and aggregate predictions from Metaculus,\footnote{\url{https://www.metaculus.com/questions/5121/date-of-artificial-general-intelligence/}} a reputation-weighted forecasting platform. These sources offer well-reasoned probability distributions for TAI development, though it's important to emphasize that such distributions are inherently speculative and represent different possibilities rather than objective truths. The source probabilities and the fitted distributions are displayed in \autoref{fig:probabilities}.

It's worth noting that these probability distributions come from individuals and communities that are particularly knowledgeable about and engaged with AI development. They likely differ substantially from the beliefs of the average household, many of whom may not be attentive to changes in the AI landscape and who expect current economic conditions to persist. This suggests an interesting extension of the model to incorporate heterogeneous beliefs, where some fraction of households maintain status quo expectations while others anticipate transformative change, possibly with varying timeline distributions. Such an extension could provide insights into how differential expectations about technological change affect wealth accumulation and economic inequality.

Using a functional form instead of directly applying the predictions from Cotra and Metaculus help to smoothen the probabilities. In addition, the negative beta binomial framework offers an intuitive way for individuals to construct their own probability distributions and see how they affect the model. By considering their beliefs about the number of necessary breakthroughs and the difficulty of achieving them, people can naturally translate their technological expectations into the model's distributional parameters.

The TFP growth rate after TAI ($g_{TAI}$) is assumed to be $30\%$ as per \citet{davidson2021could}. This is of course much higher than the $2\%$ historical average, though an increase of more than an order of magnitude is not without precedent. Prior to the industrial revolution, the world economy growth rate was near zero for most of human history \citep{roser2023gdp}.

For $\lambda$, I consider a baseline case with $\lambda = 1$ wherein AI labor is allocated proportionally to wealth. I compare with a no strategic competition case of $\lambda = 0$, and cases of more extreme competition with $\lambda = 2$ and $\lambda = 4$. The rest of the parameters are standard: $\eta = 1$, $\beta = 0.99$, $\alpha = 0.36$, $\delta = 0.025$, and $g_{SQ} = 0.018$.

\section{Equilibria and Transition Paths} \label{sec:equilibria}

This section characterizes the model's equilibrium conditions and analyzes its transition dynamics. I begin by establishing market clearing conditions and factor prices. I then examine the economy's stationary equilibria - both before TAI expectations emerge and after TAI arrives. Finally, I analyze transition paths between these equilibria, showing how strategic competition for future AI labor creates a wedge between interest rates and capital returns during the transition period.

\subsection{Market Clearing and Factor Prices} \label{subsec:markets}

In equilibrium, markets must clear and factors must be paid their marginal products. The labor, capital, and bond markets clear according to:

\begin{align*}
\int_0^1 l_i di &= L & \text{(labor)} \\
\int_0^1 k_{i,t} di &= K_t &\text{(capital)} \\
\int_0^1 b_{i,t} di &= 0 & \text{(bonds)}
\end{align*}

The bond market clearing condition reflects that bonds are in zero net supply - each household's lending must be matched by another household's borrowing. Factor prices are determined by the firm's optimization problem. The wage rate and capital rental rate are given by:

\begin{align*}
r_{k,t} &= \alpha(A_t L / K_t)^{1 - \alpha} - \delta & \text{(capital rental rate)} \\
w_t &= (1-\alpha)A_t(\alpha / (r_{k,t} + \delta))^{\alpha/(1-\alpha)} & \text{(wage)} 
\end{align*}

The interest rate on bonds is determined by no-arbitrage conditions derived from households' Euler equations for capital and bonds. In the stationary equilibrium, these equations imply that the interest rate equals the capital rental rate. However, during the transition, strategic competition for future AI labor creates a wedge between these rates. The precise determination of interest rates during the transition is detailed in \autoref{subsec:strat-wedge}.

\subsection{Stationary Equilibria} \label{subsec:stationary}

This model has two important stationary equilibria: the status-quo equilibrium before households develop beliefs regarding future TAI and the post-TAI equilibrium. Each is the standard stationary equilibrium of the neo-classical growth model with respect to their parameters, wherein the interest rate is equal to 

\begin{equation} \label{eq:stationary-interest}
    r_j = \frac{(1+g_j)^\eta}{\beta}
\end{equation}

\noindent for $j = SQ$ and $j = TAI$ respectively.

There are two important notes to make here. First, the economy will not remain in the status-quo equilibrium until the invention of TAI, but rather only until households develop beliefs regarding the invention of TAI. As soon as households realize that TAI is possible, they will begin to prepare for this uncertain future, causing the economy to exit the status-quo stationary equilibrium. The economy will then follow a transition path that will settle into a new stationary equilibrium only after TAI is invented. Alternatively, if TAI is never invented the economy will eventually settle back into the status-quo stationary equilibrium after the households believe such a transition is no longer likely.

Second, it is clear from \autoref{eq:stationary-interest} that the higher rate of post-TAI TFP growth implies a higher stationary equilibrium interest rate. While this higher post-TAI equilibrium interest rate follows mechanically from the standard neoclassical growth model's properties, the more economically interesting phenomenon is the behavior of interest rates along the transition path, particularly in the periods before TAI arrives.

These two points highlight the importance of understanding the economy's dynamic adjustment process. The transition from the status-quo equilibrium begins as soon as households develop beliefs about future TAI, and the resulting path of interest rates reflects both current economic conditions and expectations about future technological change. In the following section, I examine these transition dynamics in detail.

\subsection{Transition Paths}

The economy's transition dynamics are characterized by multiple potential paths, reflecting the uncertainty about when TAI will arrive. Once households develop beliefs about future TAI, the economy deviates from its initial equilibrium, with the path forward branching at each potential arrival date. Specifically, in each year where there is a positive probability of TAI being invented, the economy's path splits: one branch represents the scenario where TAI arrives in that year, leading eventually to the post-TAI equilibrium, while the other branch continues to the next period maintaining the possibility of future TAI arrival. This process creates S+1 distinct paths, where S is the number of years in which TAI has a positive probability of being invented: one path for TAI arriving in each possible year, plus one path where TAI never materializes and the economy eventually returns to its initial equilibrium.

Formally, a household's optimization problem before TAI reflects this branching structure. In each period, households must consider both the immediate possibility of TAI arriving and how their wealth at that moment would determine their share of future AI labor if TAI does arrive. 

I assume homogeneous initial wealth and beliefs across households, which generates identical savings decisions across households and preserves wealth homogeneity in every period. It also implies that no trade occurs in the bond market. As households will always hold zero bonds, it is sufficient to solve for the capital held by the households each period. While this symmetry assumption sacrifices realism, it permits tractable analysis of the core mechanism. Future work could relax these constraints to study wealth inequality dynamics.

The household's value function prior to TAI is:

\begin{equation}
    \begin{aligned} & V_{SQ}(k_t) = u(c_t) + \beta \left[ p_{t+1} V_{TAI}(k_{t+1}, l_i(k_{t+1}/K_{t+1})) + (1-p_{t+1}) V_{SQ}(k_{t+1}) \right] 
    \\ & \text{s.t.} \quad c_t + k_{t+1} = (1+r_{k,t})k_t + w_t l_i
    \end{aligned} 
\end{equation}

\noindent where $p_{t+1}$ is the probability of TAI arriving in the next period, and $l_i(k_{t+1}/K_{t+1})$ determines the household's share of AI labor based on their relative wealth at the time of TAI's arrival using \autoref{eq:li}.

After TAI arrives, the household's problem simplifies to:

\begin{equation}
    \begin{aligned} & V_{TAI}(k_t, l_i) = u(c_t) + \beta V_{TAI}(k_{t+1}, l_i)
    \\ & \text{s.t.} \quad c_t + k_{t+1} = (1+r_{k,t})k_t + w_t l_i
    \end{aligned} 
\end{equation}

\noindent where $l_i$ represents their fixed allocation of AI labor determined by their relative wealth when TAI arrived.

To solve for these transition paths, I employ a numerical approach based on solving for the derivatives of the value functions with respect to assets. This gradient-based method is more computationally tractable than directly solving for the value functions, as it avoids having to calculate the value function for every possible asset level. The key derivatives are:

For the pre-TAI value function:

\begin{equation} \label{eq:v-sq}
    \frac{\partial V_{SQ}(k_t)}{\partial k_{t+1}} = - u'(c_t) + \beta \left( p_{t+1} \frac{\partial V_{TAI}(k_{t+1}, f_z(k_{t+1}/K_{t+1}))}{\partial k_{t+1}} + (1-p_{t+1}) \frac{\partial V_{SQ}(k_{t+1})}{\partial k_{t+1}} \right)
\end{equation}

For the post-TAI value function:

\begin{equation} \label{eq:v-tai}
    \frac{\partial V_{TAI}(k_t, l_i)}{\partial k_{t+1}} = - u'(c_t) + \beta \frac{\partial V_{TAI}(k_{t+1}, l_i)}{\partial k_{t+1}}
\end{equation}

To find the equilibrium paths, I begin at a distant terminal period T where I assume the economy has reached its final steady state, either the post-TAI equilibrium or the initial equilibrium if TAI never arrives. Working backwards from this terminal condition, I solve for optimal household behavior in each period along each possible path.

\subsection{Interest Rate Determination and the Strategic Wedge} \label{subsec:strat-wedge}

The equilibrium interest rate is determined by households' optimal portfolio choices between bonds and capital. From the household's optimization problem, we can derive Euler equations for both assets. The Euler equation for bonds is standard:

\begin{equation} \label{eq:euler-bonds-1y}
    u'(c_t) = \beta (1+r_{b,t+1}) \mathbb{E}_{t+1}[u'(c_{t+1})],
\end{equation}

\noindent where $\mathbb{E}_{t+1}$ is the expectation operator over whether TAI will be invented next period. Though the form of the Euler equation is standard, the interest rate will differ from standard models as strategic savings will alter both current and future consumption, thereby altering current and future marginal utilities.

Expanding terms in \autoref{eq:v-sq} and simplifying, the Euler equation for capital can be expressed as:

\begin{equation} \label{eq:euler-capital}
    u'(c_t) = \beta \mathbb{E}_{t+1}[(1+r_{k,t+1}) u'(c_{t+1})] + p_{t+1} \frac{\lambda}{k_{t+1}}\sum_{i=1}^\infty \beta^i w_{t+i} u'(c_{TAI, t+i}),
\end{equation}

\noindent where $c_{TAI, t+i}$ is consumption in period $t+i$ if TAI is invented and $\mathbb{E}_{t+1}$ is the expectation operator over whether TAI will be invented next period. The second term on the right-hand side captures the additional marginal value of capital holdings due to their effect on future AI labor allocation. This term appears only in the capital Euler equation because only capital holdings, not bond holdings, influence the household's share of AI labor upon TAI's arrival.

In the standard model, the interest rate equals the marginal product of capital. However, in this model, capital ownership provides not only traditional returns but also a claim on future AI labor through the wealth-based allocation mechanism. This strategic component creates a wedge between interest rates and capital returns during the transition period: interest rates must compensate households not just for the standard opportunity cost of capital but also for giving up the strategic advantage that capital ownership provides in securing future AI labor.

The divergence between capital rental rates and interest rates emerges from the dual role of capital in this model. While both assets offer returns for postponing consumption, capital uniquely provides strategic value for securing future AI labor. This resulting wedge between rates that can be expressed as:

\begin{equation}
    r_b\mathbb{E}_t[u'(c_{t+1})] = \mathbb{E}_t[r_k u'(c_{t+1})] + \underbrace{p_{t+1}\frac{\lambda}{k_{t+1}}\sum_{i=1}^{\infty} \beta^i w_{t+i} u'(c_{TAI,t+i})}_{\text{Strategic Premium}}
\end{equation}

The strategic premium term represents the marginal value of additional capital in securing future AI labor, holding other households' savings decisions fixed. Crucially, while this premium motivates each household to accumulate more capital than they would in a standard model, in equilibrium all households increase their savings similarly. Thus, no household actually realizes the full strategic premium - instead, they reach a Nash equilibrium where each household's savings are optimal given others' identical savings choices. The elevated interest rate reflects the intensity of this competitive dynamic: households must be compensated not just for postponing consumption (captured by $r_k$) but also for the opportunity cost of not trying to ``get ahead" in the strategic competition for future AI labor, even though in equilibrium all households end up with the same share.

This strategic competition creates a form of prisoner's dilemma: all households would be better off if they could collectively agree to save at the rate implied purely by productivity growth expectations. However, the individual incentive to secure a larger share of future AI labor drives them to save more, pushing interest rates above capital rental rates even though the relative shares of AI labor remain unchanged in equilibrium.

This wedge is absent in both the pre-TAI and post-TAI stationary equilibria. In the pre-TAI equilibrium, where there are no expectations of TAI, the strategic term is zero and the two Euler equations imply $r_b = r_k$. The same holds in the post-TAI equilibrium, where AI labor shares are already determined. However, during the transition period when households anticipate TAI's possible arrival, the strategic term creates a positive wedge between these rates. The interest rate must rise above the capital rental rate to prevent households from taking infinite positions in bonds to finance capital accumulation.

The size of this wedge increases with $\lambda$, as higher values make wealth differences more pivotal in determining post-TAI outcomes. This mechanism explains why interest rates can remain elevated even as increased capital accumulation drives down the marginal product of capital. Even in periods well before TAI's potential arrival, the anticipation of wealth-based allocation of AI labor influences household saving decisions and, consequently, equilibrium interest rates.

A similar analysis applies to longer-term bonds. Consider a zero-coupon 30-year bond that pays one unit of consumption plus interest after 30 years. Its Euler equation is:

\begin{equation}
    u'(c_t) = \beta^{30} (1+r_{b,t,30})^{30} \mathbb{E}_t[u'(c_{t+30})]
\end{equation}

\noindent where $r_{b,t,30}$ is the annualized 30-year interest rate. The corresponding Euler equation for holding capital for 30 years is:

\begin{equation} \label{eq:euler-bonds-30y}
    u'(c_t) = \beta^{30} \mathbb{E}_t[(1+r_{k,t,30})^{30} u'(c_{t+30})] + \sum_{j=1}^{30} p_j \frac{\lambda}{k_{t+j}}\sum_{i=j}^{\infty} \beta^i w_{t+i} u'(c_{TAI, t+i})
\end{equation}

\noindent where $p_j$ is the probability of TAI arriving in year $j$. The second term now sums over all possible TAI arrival dates within the 30-year period. This creates a similar wedge between long-term interest rates and capital returns, though the magnitude and temporal pattern of this wedge may differ from the short-term case due to the different probability weights placed on near-term versus distant TAI arrival dates.

These theoretical insights set the stage for quantitative analysis. In the following section, I examine how different assumptions about the timing of TAI arrival and the strength of the wealth-based allocation mechanism affect key macroeconomic variables, with particular attention to saving rates and interest rates during the transition period.

\section{Results} \label{sec:result}

The model generates several key insights about how expectations of TAI affect interest rates, capital rental rates, and savings behavior. Capital rental rates are calculated using the marginal product of capital: $r_{k,t} = \alpha(A_t L / K_t)^{1 - \alpha} - \delta$.\footnote{Due to market clearing conditions, these rates must satisfy the capital Euler equation (\ref{eq:euler-capital}) in equilibrium.} One-year interest rates are derived from the household's bond Euler equation (\ref{eq:euler-bonds-1y}), while thirty-year rates are determined by equation (\ref{eq:euler-bonds-30y}). The savings rate is calculated as $(Y-C)/Y$, representing the fraction of output not consumed.

\begin{figure}[H]
    \centering
    \includegraphics[width=\textwidth]{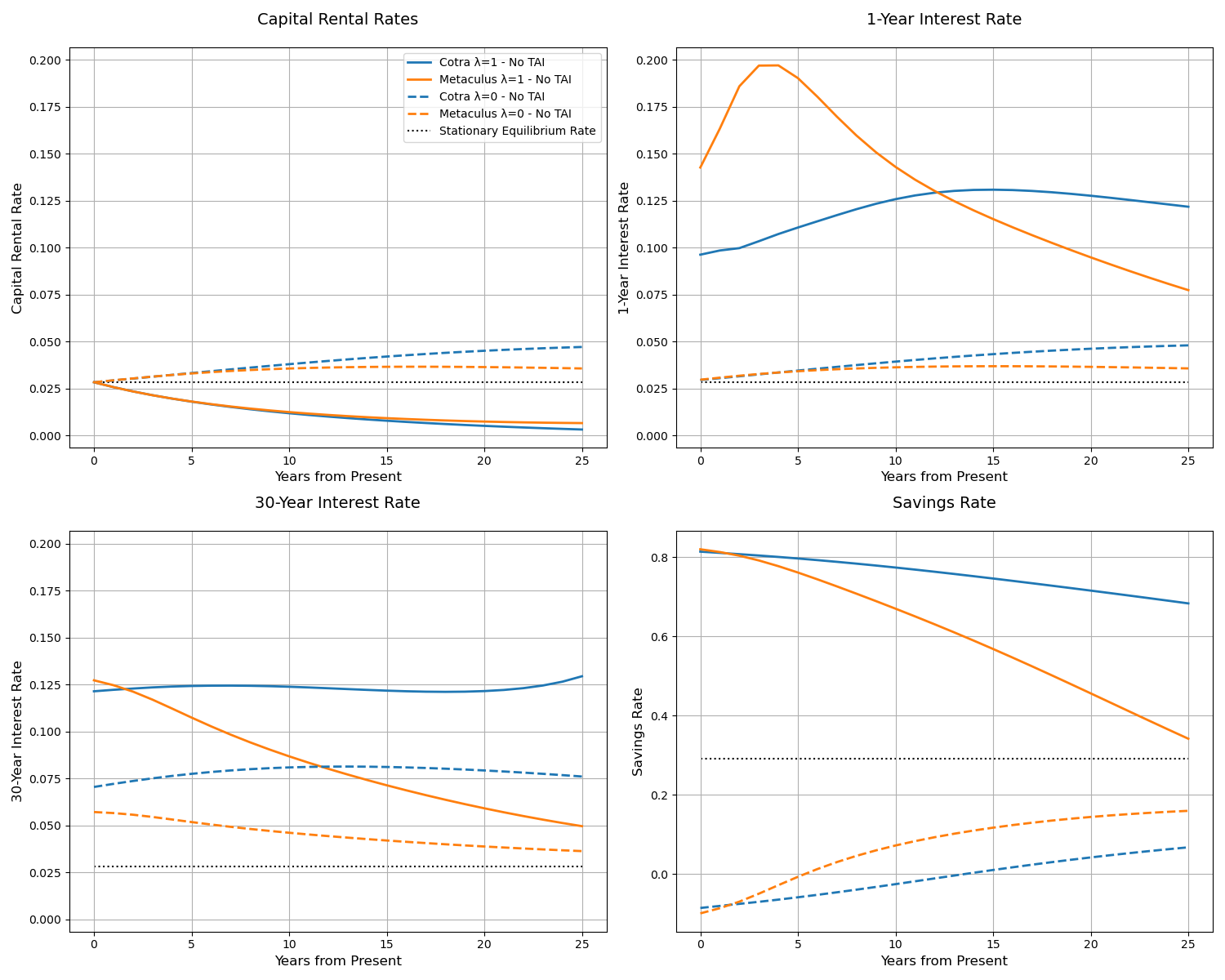}
    \caption{\small Baseline Economic Outcomes ($\lambda = 1$). The figure shows predicted paths for capital rental rates, interest rates, and savings rates under proportional wealth-based AI labor allocation. Solid lines show outcomes with strategic competition ($\lambda = 1$), while dotted lines show the no-competition benchmark ($\lambda = 0$). Blue lines represent Cotra probabilities; orange lines represent Metaculus probabilities. The horizontal dotted line shows the stationary equilibrium rate without TAI expectations.}
    \label{fig:l1}
\end{figure}

\autoref{fig:l1} presents baseline results of $\lambda=1$ under both Cotra and Metaculus probability distributions, while subsequent figures show comparative results under different values of $\lambda$, which governs the wealth-sensitivity of future AI labor allocation. All figures include the $\lambda = 0$ case (shown with dotted lines) as a benchmark that isolates pure growth expectations effects from strategic competition. The figures track pre-TAI rates—that is, rates in each year conditional on TAI not having occurred. If TAI does occur, interest rates quickly converge to a new equilibrium of approximately 35\%, as implied by \autoref{eq:stationary-interest} given the assumed post-TAI growth rate of 30\%.

\input{Figures/initial_values_table}

\subsection{Baseline Case and Strategic Competition}

Baseline simulations reveal substantial effects of TAI expectations on interest rates. As shown in Table \ref{tab:initial_values}, with moderate assumptions about wealth-based allocation of AI labor ($\lambda = 1$), one-year interest rates begin at 9.85\% under Cotra probabilities and 16.34\% under Metaculus probabilities. This represents more than a tripling of rates compared to the no-competition scenario ($\lambda = 0$) where rates are 3.05\% and 3.07\% respectively. Thirty-year rates show similar elevation: under $\lambda = 1$, they rise to 12.22\% and 12.47\% for Cotra and Metaculus probabilities respectively, compared to 7.21\% and 5.66\% without strategic competition.

The time paths of these rates, shown in \autoref{fig:l1}, exhibit distinct patterns between the probability distributions. Under Metaculus probabilities, which assign higher likelihood to near-term TAI arrival, one-year rates spike dramatically in the first five years before declining, while Cotra probabilities produce a more gradual increase followed by a modest decline. This difference reflects the more concentrated near-term probability mass in the Metaculus distribution.

The $\lambda = 0$ case, shown in dotted lines across all figures, serves as an important theoretical benchmark, isolating the pure effect of growth expectations from strategic competition effects. While growth expectations alone generate modest increases in interest rates, the much larger increases seen with positive $\lambda$ values highlight how strategic competition for AI labor control can substantially amplify these effects.

A notable feature of the model is the simultaneous presence of high savings rates and high interest rates during the transition period. As shown in \autoref{fig:l1}, savings rates begin at around 80\%. Under Cotra probabilities they remain elevated for a long time while under Metaculus probabilities savings drop more rapidly. This contrasts with household behavior without strategic behavior ($\lambda = 0$), in which savings decline under standard parameterization, and are even negative for the first 5-12 years.\footnote{Negative savings indicate that households are consuming more than current output by depleting their capital stock, reflecting their desire to front-load consumption in anticipation of higher future productivity.}\footnote{It is possible that savings could rise even with $\lambda = 0$ under smaller values of $\eta$.} This shows the important impact of strategic savings: households save aggressively to secure future AI labor allocation despite high interest rates, creating a form of prisoner's dilemma in saving behavior.

The savings rate predictions are admittedly extreme, suggesting that real-world frictions and behavioral constraints not captured in the model would likely moderate actual responses. However, the qualitative insight—that strategic competition for future AI labor can simultaneously drive up both savings and interest rates—remains relevant for understanding how TAI expectations might influence economic behavior. More realistic extensions to the model—such as incorporating heterogeneous beliefs, borrowing constraints, or gradual automation—could help generate more plausible savings behavior while preserving the core insight that strategic competition for future AI labor can simultaneously drive up both savings and interest rates. These potential extensions are discussed further in the conclusion.

\subsection{Sensitivity to Wealth-Based Allocation}

\begin{figure}[H]
    \centering
    \includegraphics[width=\textwidth]{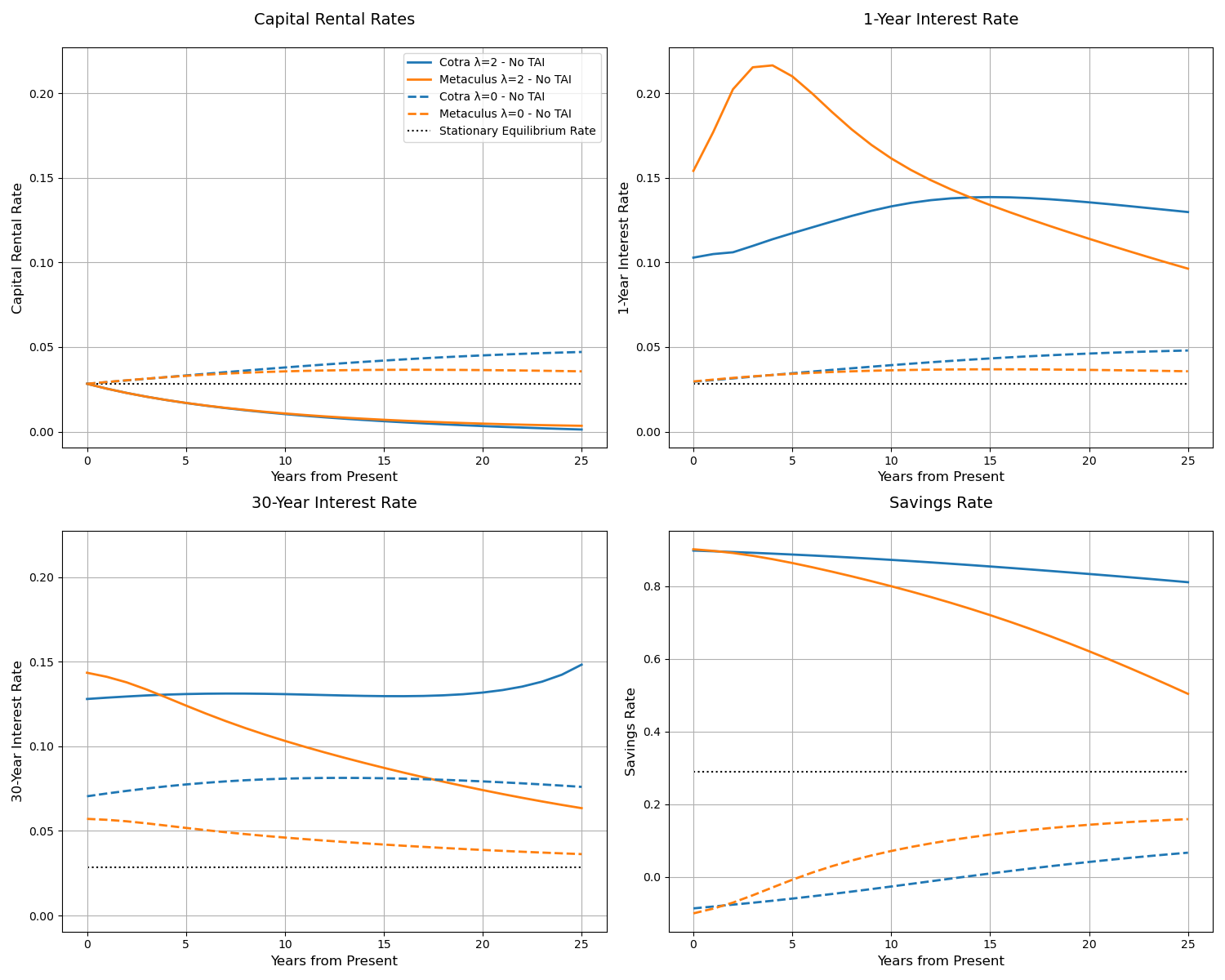}
    \caption{ \small Economic Outcomes with Enhanced Strategic Competition ($\lambda = 2$).}
    \label{fig:l2}
\end{figure}

\begin{figure}[H]
    \centering
    \includegraphics[width=\textwidth]{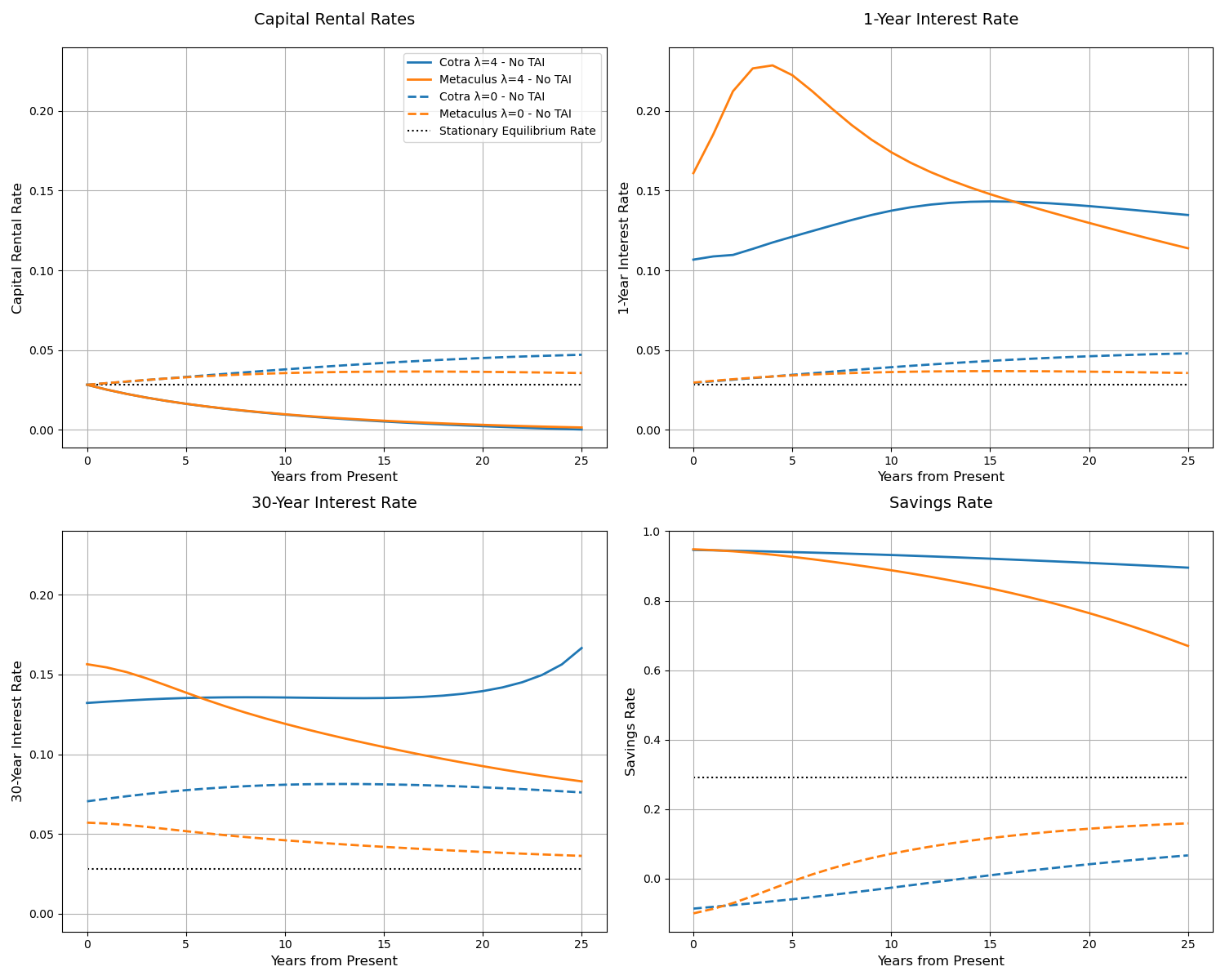}
    \caption{ \small Economic Outcomes with Strong Strategic Competition ($\lambda = 4$).}
    \label{fig:l4}
\end{figure}

Figures \ref{fig:l2} and \ref{fig:l4} explore scenarios with stronger wealth sensitivity in AI labor allocation ($\lambda = 2$ and $\lambda = 4$ respectively). As $\lambda$ increases, both short and long-term interest rates rise monotonically, reflecting intensified competition for future AI labor control. However, this effect exhibits diminishing returns: the increase in rates from $\lambda = 0$ to $\lambda = 1$ is substantially larger than subsequent increases.

For instance, under Metaculus probabilities, one-year rates increase by over 11 percentage points when moving from $\lambda = 0$ to $\lambda = 1$, but only by about 0.7 percentage points when moving from $\lambda = 2$ to $\lambda = 4$. This pattern suggests that while strategic competition for AI labor significantly affects interest rates, extreme sensitivity to wealth differences may not proportionally intensify these effects.

\subsection{Capital Rental Rates}

Capital rental rates show less dramatic variation across scenarios than interest rates and follow a generally declining pattern over time, reflecting capital accumulation. This divergence between interest rates and rental rates is particularly noteworthy: despite increasing savings driving down the marginal product of capital (as shown by falling rental rates), interest rates remain elevated due to strategic competition for future AI labor. This demonstrates how the prospect of TAI can break the traditional link between capital returns and interest rates, as households accept lower productive returns in exchange for the strategic value of wealth accumulation.

\subsection{Practical Implications}

Notably, all scenarios with positive $\lambda$ produce interest rates substantially higher than the no-competition benchmark. Even modest assumptions about wealth-based allocation of AI labor lead to dramatic increases in rates. This finding extends \citet{chow2024transformative}, whose representative agent model is equivalent to our model with $\lambda = 0$. Importantly, $\lambda$ is plausibly greater than or equal to one, as control over AI systems may concentrate disproportionately among the wealthiest actors due to their greater market power and strategic advantages in deploying AI technology. In such cases, TAI expectations could cause interest rates to rise far higher than the rates predicted by models without strategic competition, as shown in \autoref{tab:initial_values}. These findings have important implications for monetary policy and financial stability, as they suggest that evolving beliefs about TAI could create strong upward pressure on interest rates well before any technological breakthrough occurs.

\section{Conclusion} \label{sec:conclusion}

This paper develops a theoretical framework for analyzing how expectations of Transformative AI (TAI) could influence current economic behavior, with particular attention to the effect of wealth-based allocation of automated labor. The model reveals that anticipation of TAI can significantly affect present-day interest rates and capital accumulation patterns through two distinct channels: expectations of higher future growth and strategic competition for future AI labor control.

The model reveals an important game theoretic aspect of this strategic competition. While individual households are incentivized to accumulate wealth to secure larger shares of future AI labor, in equilibrium all households increase savings similarly. This creates a prisoner's dilemma where competitive wealth accumulation drives up interest rates without changing relative shares of future AI labor. This finding suggests that expectations of TAI could generate substantial financial market effects even when the strategic benefits of wealth accumulation are ultimately neutralized by equilibrium behavior.

This strategic dynamic manifests not only in elevated interest rates but also in extremely high savings rates. While such extreme savings predictions likely overstate real-world behavioral responses, they highlight how strategic competition fundamentally amplifies saving incentives. Even in standard models, higher interest rates can increase savings when the intertemporal elasticity of substitution is sufficiently high. However, our model generates substantially higher savings rates than standard calibrations through the additional channel of strategic competition, as households accept lower consumption today to compete for future AI labor control regardless of their consumption-smoothing preferences.

The results demonstrate that even moderate assumptions about wealth-based allocation of AI labor can generate substantial increases in interest rates well above predictions without strategic competition over labor control. Under baseline scenarios with proportional wealth-based allocation ($\lambda$ = 1), one-year interest rates rise to 9.85\% and 16.34\% under Cotra and Metaculus probability distributions respectively, compared to 3.05\% and 3.07\% in scenarios without strategic competition ($\lambda$ = 0). This dramatic difference highlights how competition for future AI labor can amplify saving incentives and elevate interest rates even before any technological breakthrough occurs.

A key finding is that the strength of wealth-based allocation in determining future AI labor shares (parameterized by $\lambda$) monotonically increases both short and long-term interest rates, though with diminishing returns. This suggests that while strategic competition for AI labor significantly affects financial markets, extreme sensitivity to wealth differences may not proportionally intensify these effects. Moreover, the model reveals a notable divergence between interest rates and capital rental rates, as households accept lower productive returns in exchange for the strategic value of wealth accumulation.

Several promising directions for future research emerge from this analysis. First, relaxing the assumption of homogeneous initial wealth could provide insights into how TAI expectations might affect wealth inequality dynamics. Second, incorporating heterogeneous beliefs about TAI arrival probabilities would better reflect real-world variation in technological expectations across different economic agents. Third, extending the model to include active belief updating based on observed technological progress could capture how evolving information about AI development influences economic behavior.

Additional extensions could explore the role of takeoff speed in shaping economic responses to TAI expectations. While the current model assumes an immediate transition to higher productivity growth, a more gradual takeoff might generate different patterns of anticipatory behavior. TAI could even result in superexponential growth (\citealt{aghion2017artificial}, \citealt{trammell2023economic}), which could boost interest rates even further. Furthermore, building on Chow et al.'s findings, incorporating TFP shocks could illuminate how increased growth volatility might counteract or amplify the interest rate effects identified in this paper.

These findings have important implications for both policy and theory. For policymakers, the model suggests that evolving beliefs about TAI could create significant upward pressure on interest rates well before any technological breakthrough occurs, with potential implications for monetary policy and financial stability. For economic theory, the results demonstrate how anticipation of transformative technological change and zero-sum competition can create novel strategic interactions in saving behavior, distinct from standard growth model dynamics.

In conclusion, this analysis demonstrates that expectations of TAI can substantially influence current economic behavior through both growth expectations and strategic wealth accumulation motives. As AI technology continues to advance, understanding these anticipatory channels becomes increasingly important for economic policy and planning. Future research extending this framework along the directions outlined above will be crucial for developing a more complete understanding of how technological expectations shape economic outcomes.

%\singlespacing
\setlength\bibsep{0pt}
\bibliographystyle{apalike}
\bibliography{AIbib}

% \clearpage

% \onehalfspacing

% \section*{Tables} \label{sec:tab}
% \addcontentsline{toc}{section}{Tables}

% \clearpage

% \section*{Figures} \label{sec:fig}
% \addcontentsline{toc}{section}{Figures}

%\begin{figure}[hp]
%  \centering
%  \includegraphics[width=.6\textwidth]{../fig/placeholder.pdf}
%  \caption{Placeholder}
%  \label{fig:placeholder}
%\end{figure}

% \clearpage

% \section*{Appendix A. Placeholder} \label{sec:appendixa}
% \addcontentsline{toc}{section}{Appendix A}

\end{document}

%% file: Figures/initial_values_table.tex
\begin{table}[H]
\centering
\begin{tabular}{lcc}
\hline
Parameter Set & 1y Interest Rate & 30y Interest Rate \\
\hline
\multicolumn{3}{l}{\textbf{Cotra}} \\
\quad $\lambda=0$ & 3.05\% & 7.21\% \\
\quad $\lambda=1$ & 9.85\% & 12.22\% \\
\quad $\lambda=2$ & 10.50\% & 12.88\% \\
\quad $\lambda=4$ & 10.87\% & 13.29\% \\
\hline
\multicolumn{3}{l}{\textbf{Metaculus}} \\
\quad $\lambda=0$ & 3.07\% & 5.66\% \\
\quad $\lambda=1$ & 16.34\% & 12.47\% \\
\quad $\lambda=2$ & 17.70\% & 14.11\% \\
\quad $\lambda=4$ & 18.51\% & 15.44\% \\
\hline
\end{tabular}
\caption{Interest Rates in Year 1}
\label{tab:initial_values}
\end{table}

%% file: main.bbl
\begin{thebibliography}{}

\bibitem[Aghion et~al., 2017]{aghion2017artificial}
Aghion, P., Jones, B.~F., and Jones, C.~I. (2017).
\newblock {\em Artificial intelligence and economic growth}, volume 23928.
\newblock National Bureau of Economic Research Cambridge, MA.

\bibitem[Agrawal et~al., 2018]{agrawal2018finding}
Agrawal, A., McHale, J., and Oettl, A. (2018).
\newblock Finding needles in haystacks: Artificial intelligence and recombinant
  growth.
\newblock In {\em The economics of artificial intelligence: An agenda}, pages
  149--174. University of Chicago Press.

\bibitem[Chow et~al., 2024]{chow2024transformative}
Chow, T., Halperin, B., and Mazlish, J.~Z. (2024).
\newblock Transformative ai, existential risk, and real interest rates.
\newblock Working Paper.

\bibitem[Davidson, 2021]{davidson2021could}
Davidson, T. (2021).
\newblock Could advanced ai drive explosive economic growth.
\newblock {\em Open Philanthropy}, 25.

\bibitem[Gruetzemacher and Whittlestone, 2022]{GRUETZEMACHER2022102884}
Gruetzemacher, R. and Whittlestone, J. (2022).
\newblock The transformative potential of artificial intelligence.
\newblock {\em Futures}, 135:102884.

\bibitem[Jones, 2005]{jones2005growth}
Jones, C.~I. (2005).
\newblock Growth and ideas.
\newblock In {\em Handbook of economic growth}, volume~1, pages 1063--1111.
  Elsevier.

\bibitem[Jones, 2022]{jones2022past}
Jones, C.~I. (2022).
\newblock The past and future of economic growth: A semi-endogenous
  perspective.
\newblock {\em Annual Review of Economics}, 14(1):125--152.

\bibitem[Roser et~al., 2023]{roser2023gdp}
Roser, M., Arriagada, P., Hasell, J., Ritchie, H., and {Ortiz-Ospina}, E.
  (2023).
\newblock Data page: Global {GDP} over the long run.
\newblock
  \url{https://ourworldindata.org/grapher/global-gdp-over-the-long-run}.
\newblock Part of ``Economic Growth''. Data adapted from World Bank, Bolt and
  van Zanden, Angus Maddison.

\bibitem[Trammell and Korinek, 2023]{trammell2023economic}
Trammell, P. and Korinek, A. (2023).
\newblock Economic growth under transformative ai.
\newblock Technical report, National Bureau of Economic Research.

\end{thebibliography}
